\documentclass[journal]{IEEEtran}
\newcommand{\figwidth}{3.45in}
\newcommand{\degree}{^{\circ}}

\usepackage[cmex10]{amsmath}
\usepackage{caption}
\ifCLASSINFOpdf
  \usepackage[pdftex]{graphicx}  
  \graphicspath{{./}}  
  \DeclareGraphicsExtensions{.pdf,.tif,.tiff}
\else 
  \usepackage[dvips]{graphicx}  
  \graphicspath{{./}}  
  \DeclareGraphicsExtensions{.eps,.tif,.tiff}
\fi
\usepackage{hyperref} 
\usepackage{xcolor} 
\hypersetup{
		colorlinks=true,
		linkcolor={blue!50!black},
		anchorcolor={blue!50!black},
		citecolor={blue!50!black},
		urlcolor={blue!50!black},
		pdfborder={0 0 0},
		bookmarksnumbered=true,
}
\begin{document}
\title{Optimized 3D simulation method for modeling of out-of-plane radiation in silicon photonic integrated circuits}
\author{Wouter~J.~Westerveld, 	
        H.~Paul~Urbach,        
        and~Mirvais~Yousefi% 
\thanks{
This article appeared in IEEE Journal of Quantum Electronics, vol.~47, no.~5, May 2011.
This work was supported in part by TNO.}% 
\thanks{W. J. Westerveld is with the Optics Research Group, Faculty of Applied Sciences, Delft University of Technology, 2600 GA, Delft, The Netherlands, and also with TNO, 2600 AD, Delft, The Netherlands.}%
\thanks{H. P. Urbach is with the Optics Research Group, Faculty of Applied Sciences, Delft University of Technology, 2600 AD, Delft, The Netherlands (e-mail: h.p.urbach@tudelft.nl).}%
\thanks{M. Yousefi is with TNO, 2600 AD, Delft, The Netherlands}%
\thanks{Digital Object Identifier \href{http://dx.doi.org/10.1109/JQE.2010.2099645}{DOI:10.1109/JQE.2010.2099645}}}% 
\markboth{This article appeared in IEEE Journal of Quantum Electronics, vol. 47, no. 5, May 2011 (\href{http://dx.doi.org/10.1109/JQE.2010.2099645}{DOI:10.1109/JQE.2010.2099645}).}%
{Westerveld \MakeLowercase{\textit{et al.}}: Optimized 3D simulation method for modeling of out-of-plane radiation in silicon photonic integrated circuits}
\IEEEpubid{\begin{minipage}{\textwidth}\ \\[12pt]
\begin{center}
\copyright 2011 IEEE. Personal use of this material is permitted. Permission from IEEE must be obtained for all other uses, in any current or future media, including reprinting/republishing this material for advertising or promotional purposes, creating new collective works, for resale or redistribution to servers or lists, or reuse of any copyrighted component of this work in other works.
\end{center}\end{minipage}} 
\maketitle

\begin{abstract}
We present an accurate and fast 3D simulation scheme for out-of-plane grating couplers, based on two dimensional rigorous (finite difference time domain) grating simulations, the effective index method (EIM), and the Rayleigh-Sommerfeld diffraction formula. In comparison with full 3D FDTD simulations, the rms difference in electric field is below 5\% and the difference in power flux is below 3\%. A grating coupler for coupling from a silicon-on-insulator photonic integrated circuit to an optical fiber positioned 0.1 mm above the circuit is designed as example.
\end{abstract}


\IEEEpeerreviewmaketitle

\section{Introduction}
\label{sec:intro}

\IEEEPARstart{S}{ILICON PHOTONICS} 
became one of the most promising platform for integrated photonics, due to the possibility of CMOS fabrication and thus the {\it economy of scale}, that came available in the last decade. 
While the original driver behind integrated optics was telecommunications, novel application fields such as on-chip interconnects and sensor systems have discovered the virtues of silicon photonics \cite{refs:dumon08, refs:IBM}.
Silicon photonic integrated circuits have a large refractive index contrast allowing for small device footprint, which is handy for most applications. However, it makes in-and-out coupling of light into the photonic integrated circuit (PIC) very difficult, since one has to match a $\sim$9~$\mu$m fiber core with a $\sim$0.5~$\mu$m waveguide. The use of out-of-plane grating couplers, a technique already reported in the 1970's, circumvent this problem by employing a grating that redirects the light from the waveguide upwards \cite{refs:dakss70}. Radiation occurs from an area on the top surface of the PIC, allowing for a coupler with the same dimensions as the fiber. Alignment tolerances are large, ($\sim$7~$\mu$m) for an optical fiber positioned 102~$\mu$m above the grating. Grating couplers have seen a revival due to the ascendance of CMOS fabricated silicon photonics. The large index contrast allows for broadband coupling of light ($\sim$25~nm). The gratings itself have also evolved over time from the original uniform 1D gratings to today's photonic crystal based 2D gratings \cite{refs:taillaert02,refs:taillaert04,refs:taillaert06,refs:taillaert03,refs:laere09}. 

The out-of-plane coupler is not only useful for coupling light into a fiber. Due to silicon's lack of a direct band gap, generation or detection of light requires additional material. A near-future candidate as light source for mass-produced silicon PICs is a VCSEL mounted above an out-of-plane coupler with automated pick-and-place equipment \cite{refs:thourhout08,refs:ursi10}. Furthermore, these couplers are very suitable for functional wafer-scale testing of PICs during the fabrication process. In the field of sensing, we expect grating couplers to be used for line-of-sight remote sensing in rough environments. This versatility of applications, in combination with the huge number of design parameters, requires a fast simulation method to calculate the full electromagnetic field at arbitrary  position from the grating.

Rigorous simulation methods such as finite difference time domain (FDTD) or eigenmode expansion (EMM) are required to accurately model the behavior of the high-contrast gratings. Current simulations of grating couplers are, to our knowledge, only performed as a 2D analysis (i.e. 1D gratings). Two-dimensional gratings are understood as a superposition of two 1D gratings  \cite{refs:taillaert02, refs:taillaert04, refs:taillaert06, refs:taillaert03, refs:laere09}. In this paper, we show that this two-dimensional analysis is valid only in the vicinity of the grating coupler and we extend the method with three-dimensional free-space propagation. The focus of this work is mainly on uniform gratings to demonstrate our novel calculation technique, but the analysis is easily applicable to virtually any type of out-of-plane couplers, except for couplers with two-dimensional focusing of the light.
       
The next section will detail the theory. In section~\ref{sec:simulationscheme}, we propose a novel simulation approach to describe the behavior of the grating coupler in three dimensions and compare it against full 3D FDTD simulations. In section~\ref{sec:results}, we use this method to design a uniform grating coupler within the limits of CMOS fabrication and to check the validity of full 2D simulations.

\IEEEpubidadjcol

\begin{figure}[!t]
\centering
\includegraphics[width=\figwidth]{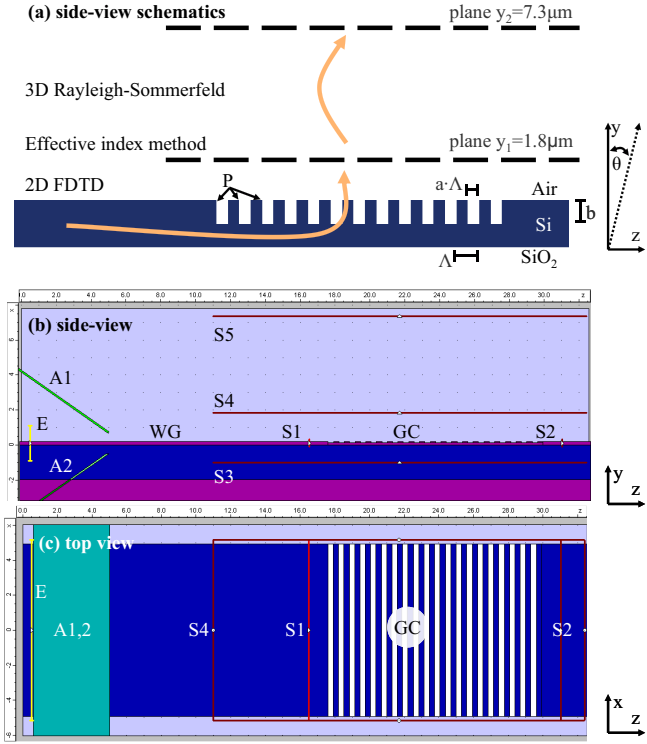} 
\caption{(a) Schematics of the simulation approach, side-view of the grating coupler. The grating period $\Lambda$, the fill factor $a$ and the etch depth $b$ are indicated. The first three "point-sources" of the intuitive description are indicated P. The planes $y_1=1.8~\mu$m and $y_2=7.3~\mu$m are indicated. The numerical methods for the different steps of the simulation approach are given.\\
(b) side-view and (c) top-view of the FDTD setup. E is the excitor. A1 and A2 are the absorbing shields. WG indicates the waveguide. GC indicates the grating coupler. S1 is the sensor for incoming and back-reflected light through the waveguide. S2 is the sensor for transmission through the waveguide after the grating coupler. S3 is the sensor downward for radiation into the substrate. S4 is the sensor for upward radiation into the air. S5 is the sensor at the position of the fiber facet (full FDTD simulation).
}
\label{fig:setup}
\end{figure}

\section{Theory}
\label{sec:theory}
An out-of-plane grating coupler couples light from a high refractive index waveguide upwards via the air into, for example, an optical fiber. Figure~\ref{fig:setup} shows a schematic of an out-of-plane grating coupler, as well as the simulation scheme we propose in this paper.

We consider the coupling of light from the fundamental mode of the input waveguide to the fundamental mode of the fiber. The simulation scheme we propose consists of four parts: a 2D FDTD simulation which describes the propagation from the waveguide to a plane just above the coupler. Then, the effective index method is applied to calculate the lateral profile of the field, based on the width of the grating in this plane, resulting in a 3D field profile. Thereafter, Rayleigh-Sommerfeld diffraction is used to propagate the field from this plane to the plane of the fiber facet and finally an overlap integral is used to calculate the coupling into the fiber mode. In what follows, we will detail the theory behind each of these steps.

The behavior of the coupler can intuitively be understood by considering all the tall-short interfaces on the left edge of the grating grooves as "point-sources" which have a phase difference dictated by the propagation speed of the light through the waveguide (a few of such "point sources" are indicated by P in Fig.~\ref{fig:setup}a). The effective (refractive) index of the grating can be estimated as the spatially weighted average of the effective indices of the fundamental modes in the tall and short parts of the waveguide. The fields emitted by these point-sources constructively interfere to a plane wave radiating under a certain angle $\theta_q$ w.r.t. the y-axis. The relation between this angle and the waveguide properties is given by \cite{refs:yariv77}:
\begin{equation} \label{eq:grating}
n_3 \sin(\theta_q) = n_\textup{eff} - q \frac{\lambda_0}{ \Lambda},
\end{equation}
where $n_3$ is the refractive index of the upper medium (air in Figure~\ref{fig:setup}), $n_\textup{eff}$ is the effective index of the grating, $q$ is the coupling order, $\lambda_0$ is the wavelength of the light in vacuum and $\Lambda$ is the grating period. For perfect vertical coupling, $\theta_q=0\degree$, Eq.~(\ref{eq:grating}) describes the second order resonance of a distributed Bragg reflector (DBR) \cite{refs:yariv77}, which very efficiently reflects the forward propagating light in the waveguide backwards, rather than radiating it upwards. Equation~(\ref{eq:grating}) can also be derived by treating the grating as a perturbation of the waveguide and calculating the coupling coefficient between the waveguide mode and a plane wave radiating with angle $\theta_\textup{q}$. The coupling strength scales with the fill factor $a$ as $\sin(\pi a)$ and depends linearly on the etch depth $b$ \cite{refs:streifer76}. For efficient coupling to a fiber, the grating should be designed such that only one coupling order ($q=1$) is allowed, and the coupling strength should be designed between strong upwards coupling and low backwards reflections. The ratio between upwards and downwards radiation is strongly influenced by the buried oxide (BOX) height. For higher coupling efficiency, a non-uniform grating can be designed \cite{refs:taillaert04,refs:taillaert06}.

\subsection{Two-dimensional calculations and the effective index method} 
\label{sec:theeim}

The width of the vertical grating coupler is much larger than its height, thus the variation of the electromagnetic fields of the fundamental mode of the coupler is slow in the x-direction compared to its variation in the (y,z)-plane. A two-dimensional (y,z) analysis can therefore be used in the vicinity of the coupler up to approximately one wavelength above the grating. To obtain the lateral profile in a plane just above the coupler, we apply a method similar to the effective index method (EIM) by approximating the field $\bf{E}$ as ${\bf E}(x,y,z) = {\bf E}^{2D}(y,z) \cdot E_x^{mode}(x)$, where ${\bf E}^{2D}$ is the electric field as calculated using two-dimensional analysis (assuming invariance in the x-direction) and lateral field profile $E_x^{mode}(x)$ is approximated from the lowest order mode in the waveguide.

\subsection{Propagation into the upper-half space}
\label{sec:thers}

The Rayleigh-Sommerfeld diffraction equation describes the electromagnetic field in a semi-infinite homogeneous medium, irradiated from a finite aperture \cite{refs:goodmanfo}. In our case, the finite aperture is a sufficiently large part of the plane just above the grating coupler and air is the homogeneous medium. The field outside this aperture is assumed to be zero. The diffraction equation for monochromatic light (vacuum wavelength $\lambda_0=2\pi/k_0$) with free-space propagation constant $k=k_0n_3$ is written as \cite{refs:goodmanfo}:
\setlength{\arraycolsep}{2pt}
\begin{eqnarray} \label{eq:RS}
U(x,y,z) &=& \iint \limits_{\textup{aperture}} U(x',y_0',z') G(x,y,z; x',y_0',z') dx'dz', \nonumber \\
\textup{where} \quad\;\;\; & & \\
G & = & \frac{1}{2\pi} (1- \imath kr) \frac{(y-y_0')}{r} \frac{\exp(\imath kr)}{r^2}, \\
r & = & \sqrt{(x-x')^2 + (y-y_0')^2 + (z-z')^2},
\end{eqnarray}
\setlength{\arraycolsep}{5pt}
and U is any electric or magnetic field component in phasor notation with time dependence given by $e^{-i\omega t}$. Greens' function $G$ is the sum  of the fields of two in phase point sources that are images of each other with the plane $y=y_0'$ as mirror. This choice of Greens' function allows the field in the air to be described as a function of the field in the aperture, not requiring knowledge of the divergence of the field. When the field is calculated for a horizontal plane (y constant), $G$ depends on $x-x'$ and $z-z'$, so Eq.~(\ref{eq:RS}) is a 2D convolution, which can be very efficiently calculated using fast-fourier-transforms (FFTs) \cite{refs:shen06}. When the plane is tilted along the x-direction, the inner integral over $x'$ is a convolution and can be calculated using FFTs. 
The 2D equivalent of Eq.~(\ref{eq:RS}), where $U=U(y,z)$, is obtained by integrating Eq.~(\ref{eq:RS}) along $x'$, resulting in \cite{refs:berkhout}:
\setlength{\arraycolsep}{2pt}
\begin{eqnarray} \label{eq:RS2D}
U(y,z) & = &\int \limits_{\textup{aperture}} U(y_0',z') G(y,z; y_0',z') dz', \quad\quad\quad\quad\quad\\
\textup{where} \quad\quad\; & & \nonumber\\
G & = & \frac{\imath k}{2} \frac{(y-y_0')}{r} H^{(1)}_1(k r), \\
r & = & \sqrt{(y-y_0')^2 + (z-z')^2}, 
\end{eqnarray}
\setlength{\arraycolsep}{5pt}
where $H^{(1)}_1(k r)$ is the first-order Hankel function of the first kind,
 i.e. $H^{(1)}_1(k r) = J_1(kr)+\imath Y_1(kr)$ where $J_1$ and $Y_1$ are the first order Bessel functions of the first and second kind, respectively.

\subsection{Coupling into a fiber}
\label{sec:thefiber}

For monochromatic waves in dielectrics, the time-averaged power flux $\Phi$ through a surface can be calculated as \cite{refs:hunsperger}:
\begin{equation} \label{eq:flux}
\Phi = \frac{1}{2} \int_{\textup{surface}}{\textup{Re}\left( {\bf{E} \times \frac{1}{\mu} \bf{B}^{\star}} \right ) d\bf{S}},
\end{equation}
where $\bf{E}$ and $\bf{B}$ are the phasor notations of the electric and magnetic fields, respectively. 
The power coupling efficiency into a fiber mode with electric field ${\bf{E}}^f(x,\rho)$ is estimated as \cite{refs:hunsperger}:
\begin{equation} \label{eq:eta}
\eta_\textup{overlap} = \frac{| \iint  E_x^i(x, \rho) \cdot E_x^f(x, \rho) d{\bf{S}} |^2} 
               					{\iint{| E_x^i(x, \rho)|^2 d\bf{S} } \cdot \iint{ | E_x^f(x, \rho) | ^2 d{\bf{S}} }}
               					,
\end{equation}
where ${\bf{E}}^i$ is the electric field incident on the fiber facet. Coordinates $x$ and $\rho$ are in the tilted plane, parallel to the fiber facet. The fiber mode is approximated as a Gaussian beam with a beam diameter $2 w_0 = 10.4$~$\mu$m. Details of coupling from air into the fiber are neglected since there is a small refractive index step and a small angle of incidence of the incoming wave. When the field $E_x^i$ can be separated in $x$ and $\rho$ dependence, i.e  $E_x^i(x, \rho) = E_x^{i,x}(x) \cdot E_x^{i,\rho}(\rho)$, then the overlap $\eta_\textup{overlap}$ can also be separated, i.e. $\eta_\textup{overlap} = \eta_{\textup{overlap},x} \cdot \eta_{\textup{overlap},\rho}$. 

%\IEEEpubidadjcol

\begin{table}[!t]
\renewcommand{\arraystretch}{1.3}
\caption{Simulation speeds and memory requirements for the different numerical methods as presented in Sec.~\ref{sec:simulationscheme}. All simulations are performed with a 44~nm numerical gridsize. Note that the simulations in Sec.~\ref{sec:results} have a numerical gridsize of 20~nm or 10~nm, when necessary. Rayleigh-Sommerfeld (R.-S.) calculations are implemented in \textsc{Matlab}. This test is performed on 32bit PC with Windows XP professional, an Intel Xeon CPU 5130 @ 2.00 GHz and 3.00GB of RAM.}
\label{tbl:speed}
\centering
\begin{tabular}{c||c||c||c}
\hline
\bfseries Method & \bfseries Window [$\mu$m]& \bfseries Time & \bfseries Memory \\
\hline\hline
3D FDTD (upto $y_2$)					& 32.56 x 12.10 x  11 			& 17h & 2.62GB \\ \hline
3D FDTD (upto $y_1$)					& 32.56 x 12.10 x 7.7				&12h	& 2.15GB \\ \hline
2D FDTD (upto $y_2$)					& 32.56 x 11								& 2m	& 80MB \\ \hline
2D FDTD (upto $y_1$)					& 32.56 x 7									& 1m	& 20MB \\ \hline
3D R.-S. (horizontal plane)		& surface 21.42 x 10.34			& 2s	& $<$100MB \\ \hline
3D R.-S. (tilted plane) 			&	surface 21.42 x 10.34			&	10m &	$<$100MB \\ \hline
2D R.-S. 											& length 21.42							& 1s 	& $<$1MB \\ \hline
Effective Index Method 				& surface 21.42 x 10.34			& 1s	& $<$11MB \\ \hline
\hline
\end{tabular}
\end{table}

\begin{figure}[!t]
\centering
\includegraphics[width=\figwidth]{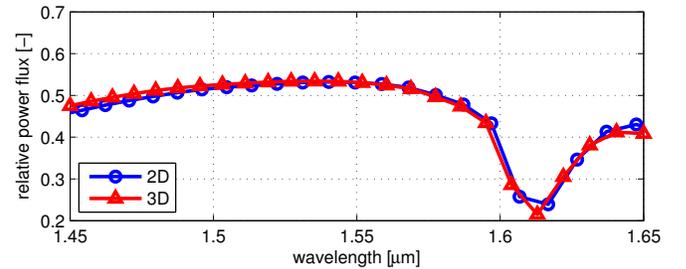} 
\caption{Flux versus wavelength. Comparison of 3D FDTD simulations and 2D FDTD simulations.}
\label{fig:flux2d3d}
\end{figure}

\begin{figure}[!t]
\centering
\includegraphics[width=\figwidth]{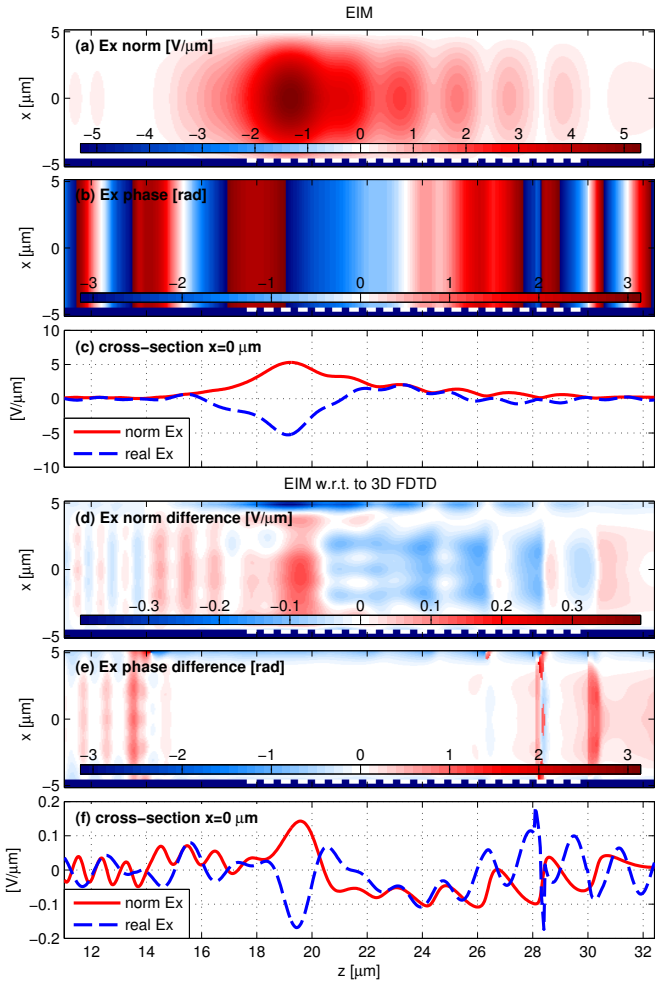} 
\caption{Effective index method (EIM) compared to full 3D FDTD simulations. The plots present the $E_x$ field in the $y_1$-plane. Plots (a-c): $E_x$ field as calculated using the EIM. Plots (d-f): difference between EIM and 3D FDTD simulations, i.e. $E_x^{\textup{EIM}}-E_x^{\textup{3D FDTD}}$. Plots (a,b,d,e) show a top-view with the z-direction and the x-direction along the horizontal and vertical axes, respectively. The position of the grating coupler is indicated in the bottom of each plot. Plots (c and f) show a cross-section through $x=0$.}
\label{fig:eim}
\end{figure}

\section{Simulation scheme}
\label{sec:simulationscheme}

We suggest a novel combination of numerical methods, see Fig.~\ref{fig:setup}, that employs the theory described in the previous section. The electromagnetic field up to plane just above the coupler is calculated using 2D FDTD simulations and the lateral profile is obtained using the effective index method (EIM). Propagation from this plane upwards is performed using the 3D Rayleigh-Sommerfeld diffraction formula. The method, as well as individual steps, are compared against full 3D FDTD simulations. Table 1 lists the calculation speeds of the different steps detailed in this section.

The grating coupler we investigate in this section has a silicon-on-insulator (SOI) layer stack which consists of a silicon substrate, covered by a 1980~nm thick SiO$_2$ layer for optical isolation between the substrate the 220~nm top mono-crystalline silicon device layer. The grating is partially etched into this device layer (see Fig.~\ref{fig:setup}b). The coupler waveguide has width $W=9.9$~$\mu$m in the x-direction, the grating has period $\Lambda=616$~nm, etch depth $b=88$~nm, fill factor $a=50$\% and $N=20$ fundamental periods. The first groove of the grating is at z~=~17.6~$\mu$m. FDTD simulations are performed in CrystalWave \cite{refs:photondesign} with a 44~nm gridsize, which is a bit large for this grating but does describe all the relevant phenomena. A mode excitor launches a light pulse with central wavelength $\lambda_0=1.55$~$\mu$m and bandwidth 0.2~$\mu$m, forward traveling from z~=~0.5~$\mu$m in the fundamental TE mode of the waveguide. In order to avoid artifacts of the excitor, we introduce two strongly absorbing and reflecting shields (A1 and A2 in Fig.~\ref{fig:setup}b) in the system, such that at the point indicated WG in Fig.~\ref{fig:setup}b, all the power resides in the input waveguide. The power flux is normalized to the power flux incident on the grating coupler, as measured by sensor 1 (S1).

The $x$ component of the electric field, $E_x$, is much larger than the other components (3 orders of magnitude) and its real part is therefore used to measure the difference between alternative simulation methods ($E_x$) and a FDTD simulations ($E_x^{\textup{FDTD}}$). Both phase and amplitude are taken into account. For the field in a surface, the root-mean-square difference is used as follows:
\begin{equation} \label{eq:deltae}
\Delta E_x \equiv \sqrt{ \frac{
		\iint \limits_{\textup{surface}} \textup{Re}\{E_x - E_x^{\textup{FDTD}}\}^2 dx dz 
		}{ 
		\iint \limits_{\textup{surface}} \textup{Re}\{E_x^{\textup{FDTD} }\}^2 dx dz
		} }.
\end{equation}
Another measure used is the relative difference in power flux $\Phi$. The latter takes all electric and magnetic field components into account.

\subsubsection{The effective index method}

Two-dimensional simulations in the (y,z)-plane are used in the vicinity of the coupler up to a plane at $y_1 \equiv 1.8$~$\mu$m above the grating, see Fig.~\ref{fig:setup} (b) and (c). The power flux through this plane is shown versus wavelength in Fig.~\ref{fig:flux2d3d} for both 2D and 3D simulations. The DBR behavior of the coupler is clearly visible at $\lambda =1.61~\mu$m as expected from Eq.~(\ref{eq:grating}). The difference between 2D and 3D simulations at central wavelength of 1.55~$\mu$m is below 1\%.

\begin{figure}[!t]
\centering
\includegraphics[width=\figwidth]{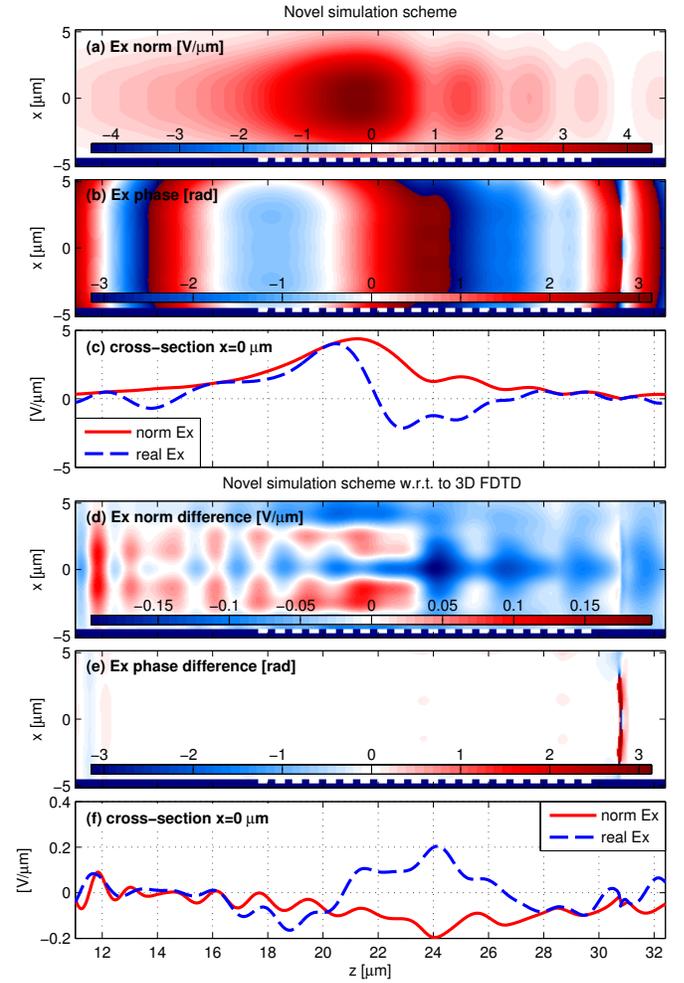} 
\caption{Our novel simulation approach compared to full 3D FDTD simulations. The plots present the $E_x$ field in the $y_2$-plane. Figure layout is identical to Fig.~\ref{fig:eim}}
\label{fig:novsim}
\end{figure}

To obtain the lateral profile in the plane just above the coupler, we apply a method similar to the effective index method (EIM). The lateral field profile $E_x^{mode}(x)$ is approximated from the lowest order mode in the waveguide which is calculated using the film mode matching (FMM) mode solver as implemented in the FimmWave \cite{refs:photondesign} software. Figure~\ref{fig:eim} shows that the difference between the EIM (which uses 2D FDTD) and the full 3D FDTD, is $\Delta E_x = 5$\%.

\subsubsection{Rayleigh-Sommerfeld diffraction formula}

Using the field at $y_1$ as obtained from a 3D FDTD simulation, we calculate the field at $y_2 \equiv 7.3$~$\mu$m using the Rayleigh-Sommerfeld diffraction formula. We find that the difference in field as compared to a full 3D FDTD simulation, $\Delta E_x=3$\%, and the difference in power flux, $\Delta \Phi=2$\%. For full two-dimensional calculations using the 2D Rayleigh-Sommerfeld diffraction formula, Eq.~(\ref{eq:RS2D}), the differences with FDTD are $\Delta E_x=2$\% and $\Delta \Phi=1$\%.

\subsubsection{Novel simulation scheme}

The results of our simulation approach are presented in Fig.~\ref{fig:novsim}. The difference in electric field is $\Delta E_x = 5$\% and the difference in the flux is $\Delta \Phi=3$\%. We therefore conclude that this approach is accurate enough to use in optimizing the design of grating couplers which couples light to fibers or to other optical components at arbitrary distance from the grating coupler.

\begin{figure}[tbp]
\centering
\includegraphics[width=\figwidth]{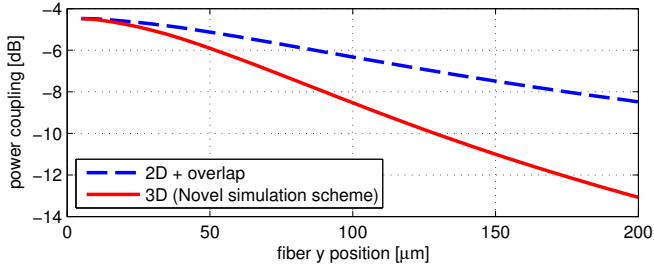} 
\caption{Power coupling between the fundamental waveguide-mode and the fiber-mode. The distance of the PIC to the fiber (e.g. the fiber y-position) is plotted along the horizontal axis. For each y-position, the z-position and tilt of the fiber are optimized for maximum coupling. Grating parameters as specified in Sec.~\ref{sec:simulationscheme}. This result is calculated using our novel method presented in Sec.~\ref{sec:simulationscheme}. }
\label{fig:fibercoupling2d3d}
\end{figure}

\section{Results}
\label{sec:results}

This section presents results we obtained with our 3D simulations of out-of-plane grating couplers. In Sec.~\ref{sec:res2d3d}, we investigate up to which height the full 2D theory is valid. In Sec.~\ref{sec:design}, as a demonstration of our method, we design a grating which is optimized for coupling to an optical fiber that is positioned 102~$\mu$m above the chip. 

\subsection{2D versus 3D simulations}

\label{sec:res2d3d}
Fully two-dimensional simulation approaches can be used to calculate fields in the vicinity of the grating coupler. The coupling efficiency is then calculated by applying the effective index method on the fiber facet, which allows to correct the 2D simulation with a simple overlap efficiency in the lateral direction, $\eta_{\textup{overlap},x}$, as shown in Sec.~\ref{sec:thefiber}. Further away from the grating, the field will spread in lateral direction and a 3D approach is required. In Fig.~\ref{fig:fibercoupling2d3d}, we show the difference between 3D and 2D simulations. At each fiber height, the fiber $z$-position (longitudinal position) and tilt $\theta$ are optimized for maximum coupling. 
The difference in to-fiber power coupling between a fully 2D analysis and a 3D analysis is $\sim$2~dB when the fiber is 100~$\mu$m above the grating.

\begin{figure}[!t]
\centering
\includegraphics[width=\figwidth]{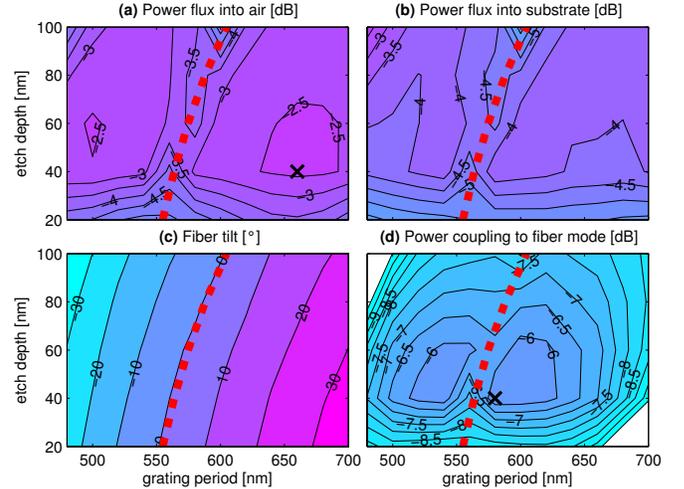} 
\caption{Optimizing of the grating parameters. (a) Power radiated into the air. (b) Power radiated into the substrate. (c) Fiber tilt for maximum power coupling to the fiber. (d) Power coupling to the fiber. Result is calculated using our novel simulation approach. The grating parameters ($\Lambda$ and $b$) for maximum upwards coupling and for maximum coupling to the fiber are indicated X. The dashed line indicates the grating parameters where a Bragg reflection is expected. The grating has a fill factor $a=50$\% and width $W=22$~$\mu$m.}
\label{fig:scanLb}
\end{figure}

\subsection{Grating coupler design}
\label{sec:design}

This section presents design steps for an out-of-plane grating coupler which couples light to a single mode fiber positioned 102~$\mu$m above the chip. In order to allow for mass-fabrication, we design a grating coupler that obey's the latest design rules of industrial CMOS lines. The EU-FP7 consortium {\it ePIXfab}, see \cite{refs:dumon09}, offers access to high-end photonic IC technologies at IMEC and LETI and we will use their design rules \cite{refs:letiflex5}, which require a 2~$\mu$m thick BOX layer and a minimal feature (groove or line) size of 120~nm. The height of our waveguide is 220~nm to suppress TM-modes. We found that the most dominant parameters in the grating design are the grating period $\Lambda$ and the etch depth $b$. We will therefore start by optimizing the grating using these parameters and later fine-tune the system suing the other parameters as mentioned in Sec.~\ref{sec:intro}.

The FDTD simulations have a numerical gridsize of 20~nm or, when required, 10~nm and is chosen such that it always perfectly represents the material profile of the grating coupler. Free-space propagation has a numerical gridsize of 40~nm. The absorbing/reflecting shield is omitted as the numerical artifact of direct air radiation from the excitor does not influence the field at the fiber facet using this simulation scheme. For each set of grating parameters, the fiber position and tilt are optimized using a fully 2D approach. The first guess of the fiber tilt for this optimization is given by Eq.~(\ref{eq:RS}) and the first guess of fiber z-position is the maximum $E_x$ field component in the plane 102~$\mu$m above the coupler. The optimization steps are 100~nm in position and 0.1$\degree$ in tilt. After finding the optimal fiber position, the coupling efficiency is calculated using a 3D simulation. 

\begin{figure}[!t]
\centering
\includegraphics[width=\figwidth]{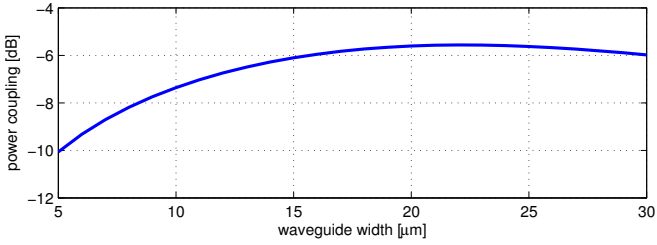} 
\caption{Power coupling between the fundamental waveguide mode and the fiber mode. The width of the waveguide is plotted along the horizontal axis.}
\label{fig:scanwidth}
\end{figure}

\begin{figure}[!t]
\centering
\includegraphics[width=\figwidth]{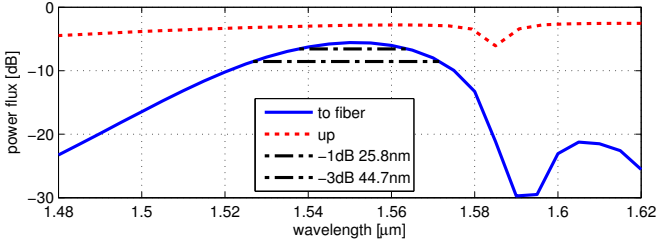} 
\caption{Bandwidth of the grating coupler with parameters ($\Lambda$~=~580~nm, $b$~=~40~nm, $a$~=~50\%).
}
\label{fig:vswavelength}
\end{figure}

A parameter scan of $\Lambda$ and $b$  in steps of 20~nm is presented in Fig.~\ref{fig:scanLb}, with the fill factor $a$~=~50\% and the width $W$~=~22~$\mu$m kept constant. The grating always has $N$~=~60 periods because there is only very little light left in the waveguide after this number of periods due to the strong coupling. Figures~\ref{fig:scanLb} (a) and (b) show the amount of upwards and downwards coupled light, which is calculated using the 2D FDTD simulation (also see Fig.~\ref{fig:setup}). They show the same trend in dependence on $\Lambda$ and $b$, which makes it impossible to discriminate against downwards coupling without extra tricks. The dotted line is the position where the grating is expected to behave as a Bragg reflector, from Eq.~(\ref{eq:grating}) with $\theta_1=0\degree$. The dip in both up- and downwards flux is clearly observed. Figure~\ref{fig:scanLb}c shows the fiber tilt for optimal coupling into the fiber, this tilt is equal to the angle of the dominant plane-wave in the radiated field because the overlap of the fiber mode with the incident field is highest for a constant phase front. The prediction of the DBR line accurately describes vertical coupling. The difference between the optimal fiber tilt and the out-coupling angle $\theta_1$ as predicted by Eq.~(\ref{eq:grating}) was below $3.5\degree$ for all simulations in this plot. The total coupling efficiency from the waveguide to the fiber is presented in Figure~\ref{fig:scanLb}d. One sees that the grating parameters for maximum upwards coupling and maximum coupling to the fiber are not the same. In Fig.~\ref{fig:scanLb}a, the maximal coupling is observed for period $\Lambda$~=~580~nm and etch depth $b$~=~40~nm. From this parameters, we optimized the duty cycle $a$, which was varied from 120~nm to 480~nm in steps of 20~nm. The coupling efficiency to the fiber showed only a weak change when varying $a$, except when the effective index of the grating was such that the DBR reflection showed up. The maximum was found at $a = 50$\%. 

\begin{figure}[!t]
\centering
\includegraphics[width=\figwidth]{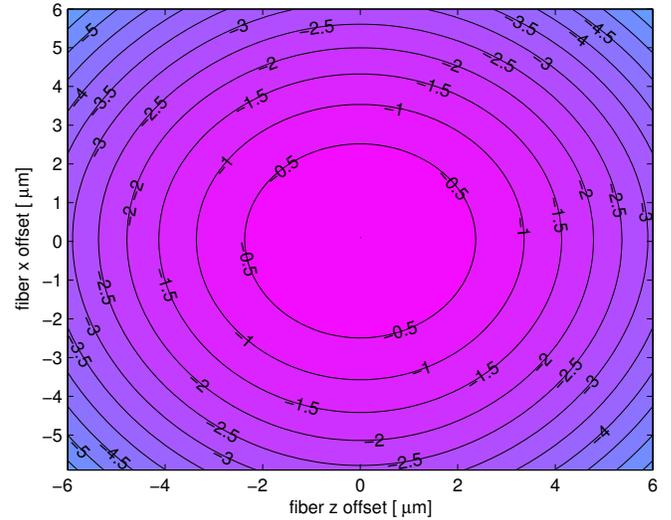} 
\caption{Power coupling from the waveguide to the fiber for fiber alignment offsets as indicated by the axes. Grating coupler with parameters ($\Lambda$~=~580~nm, $b$~=~40~nm, $a$~=~50\%). 
}
\label{fig:alignment}
\end{figure}

The next parameter being optimized is the width of the grating coupler waveguide, as presented in Fig.~\ref{fig:scanwidth}. The optimal width is found to be 22~$\mu$m. 

For this coupler, the bandwidth for a maximum transmission loss of 1~dB is 25~nm, as shown in Fig.~\ref{fig:vswavelength}. The fiber is 102~$\mu$m above the coupler, so light of a different wavelength will radiate under a different angle, thereby not fully aiming at the fiber. The fiber position was given offsets in z- and x-directions to test the alignment tolerances. The area with a maximum transmission loss of 1dB has a diameter of 6.8~$\mu$m and is almost circular (see Fig.~\ref{fig:alignment}).

\begin{figure}[!t]
\vspace{-5mm}
\centering
\includegraphics[width=\figwidth]{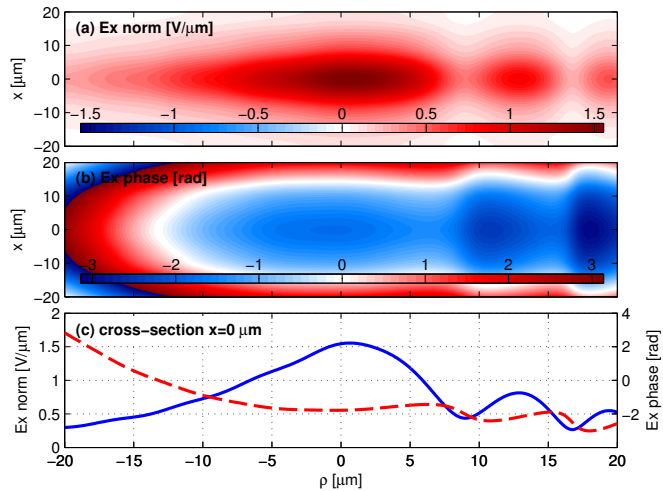} 
\caption{Electric field (x component) at the plane of the fiber facet. (a) and (b) are a top-view with the $\rho$ and the x along the horizontal and vertical axis, respectively. Plot (a) shows the amplitude of $E_x$. Plot (b) shows the phase of $E_x$. The overall phase of the field is rotated for clarity of the plot. Plot (c) is a cross-section of the field at $x$~=~0~$\mu$m. The solid and dashed line are the amplitude (left axis) and phase (right axis), respectively. The field originates from a grating with parameters ($\Lambda$~=~580~nm, $b$~=~40~nm, $a$~=~50\%).
}
\label{fig:fieldb40}
\end{figure}

\begin{figure}[!t]
\vspace{-5mm}
\centering
\includegraphics[width=\figwidth]{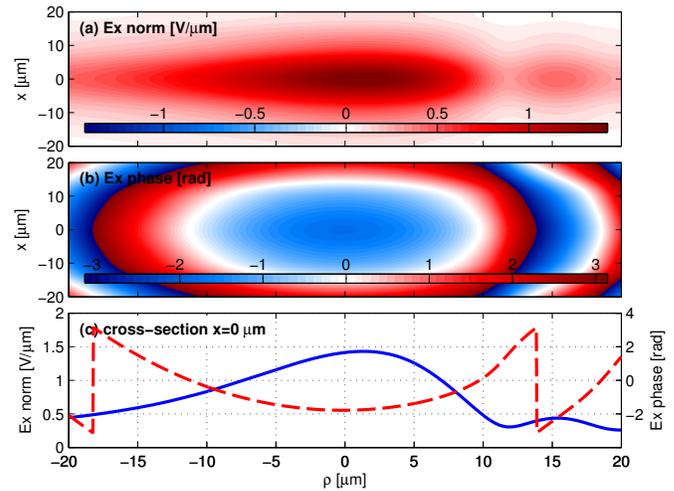} 
\caption{Figure layout is identical to Fig.~\ref{fig:fieldb40}. This field originates from a grating with parameters ($\Lambda$~=~630~nm, $b$~=~70~nm, $a$~=~50\%).
}
\label{fig:fieldb70}
\end{figure}

The optimal grating parameters we found are slightly different from the ones in Ref. \cite{refs:taillaert06}. This can be attributed to the fact that we chose the fiber to be positioned further away from the grating. Our etch depth is shallower, resulting in less strong coupling per propagated length through the coupler, and thus a larger radiative area. This gives rise to less divergence of the electromagnetic field when propagating upwards. To illustrate this, Figures \ref{fig:fieldb40} and \ref{fig:fieldb70} present the field at the plane of the fiber facet. The first figure shows the plane as irradiated by a grating coupler with a shallow etch depth, $b$~=~40~nm. The second grating coupler as used for the second figure has a deeper etch depth, $b$~=~70~nm. In both cases, the position of the fiber is optimized for maximum coupling, as in Fig.~\ref{fig:scanLb}. The figures show that the field is slightly less diverged for the shallow grating and also that the phase of the field is much more constant over the positions where the amplitude of the field is strongest. The novel simulation method gives the opportunity to inspect these fields and do this analysis.

\section{Conclusion}
In this paper, we presented a novel three-dimensional simulation scheme and used it to simulate out-of-plane grating couplers. The method is fast and accurate enough to meet the requirements for grating coupler design, involving a huge number of design parameters. In comparison with 3D FDTD simulations, our method showed a root-mean-square difference in electric field below 5\% and a difference in power flux below 3\%, while improving computational speed by two orders of magnitude. 

We foresee a broad range of applications for grating couplers in silicon photonics, and our method fulfills this versatile need. The presented method is applicable to may out-of-plane radiation simulations, such as coupling to the emission of a VCSEL, or on-wafer inspection of photonic integrated circuits. 
  
Current simulations of grating couplers are, to our knowledge, only performed as a 2D analysis. We found that this approximation is only valid in the vicinity of the coupler by comparing 2D and 3D FDTD simulations. 

As an example of our method, we designed a CMOS compatible grating coupler for coupling of light from a waveguide to a fiber at a distance of 0.1 mm from the photonic integrated circuit. 

In this work, a FDTD method is used to calculate the electromagnetic fields in the vicinity of the grating coupler, however, this is not a limitation. Other methods such as the eigemode expansion technique (EME) can also be used in the presented scheme \cite{refs:taillaert02}. In the field of sensing, we expect line-of-sight remote readout of silicon photonic integrated circuits to become possible using grating couplers, opening the possibility of sensing in extreme environments without need for a physical interconnect such as a fiber.

% \bibliographystyle{IEEEtran}
% \bibliography{IEEEabrv,refs}

\end{document}